\documentclass[trackchanges]{aastex631}
\usepackage{graphicx}
\usepackage{amsmath}
\usepackage{multirow}

\renewcommand{\i}{\mathrm{i}}
\shorttitle{Mercury Instability Rare Event Sampling}
\shortauthors{Abbot et al.}

\begin{document}

\title{Rare Event Sampling Improves Mercury Instability Statistics}

\correspondingauthor{Dorian S. Abbot}
\email{abbot@uchicago.edu}

\author{Dorian S. Abbot}
\affiliation{Department of the Geophysical Sciences \\
The University of Chicago \\
Chicago, IL 60637 USA}

\author{Robert J. Webber}
\affiliation{Courant Institute of Mathematical Sciences \\
New York University \\
New York, NY 10012 USA}

\author{Sam Hadden}
\affiliation{Harvard-Smithsonian Center for Astrophysics \\
The Institute for Theory and Computation\\
Cambridge, MA 02138, USA}

\author{Darryl Seligman}
\affiliation{Department of the Geophysical Sciences \\
The University of Chicago \\
Chicago, IL 60637 USA}

\author{Jonathan Weare}
\affiliation{Courant Institute of Mathematical Sciences \\
New York University \\
New York, NY 10012 USA}

\begin{abstract}
Due to the chaotic nature of planetary dynamics, there is a non-zero probability that Mercury's orbit will become unstable in the future. Previous efforts have estimated the probability of this happening between 3 and 5 billion years in the future using a large number of direct numerical simulations with an N-body code, but were not able to obtain accurate estimates before 3 billion years in the future because Mercury instability events are too rare. In this paper we use a new rare event sampling technique, Quantile Diffusion Monte Carlo (QDMC), to estimate that the probability of a Mercury instability event in the next 2 billion years is approximately $10^{-4}$ in the REBOUND N-body code. We show that QDMC provides unbiased probability estimates at a computational cost of up to 100 times less than direct numerical simulation. QDMC is easy to implement and could be applied to many problems in planetary dynamics in which it is necessary to estimate the probability of a rare event.
\end{abstract}

\section{Introduction} \label{sec:intro}

\citet{laskar1994large} used secular simulations, with the equations of motion averaged over planetary orbits, to reach the surprising conclusion that it is possible for Mercury's orbit to become unstable. The mechanism underlying Mercury's potential instability involves resonances among the secular system's modes of oscillation that can transfer angular momentum among the planets and dramatically increase Mercury's eccentricity \citep{LithwickWu2011secular_chaos,Boue2012,lithwick2014secular,batygin2015chaotic}. As a result, Mercury can pass very near Venus, disrupting its orbit, and leading to a collision with the Sun or another planet. \citet{laskar2008chaotic} performed a large number of secular simulations with slightly different initial conditions to estimate that the probability of this happening is $\mathcal{O}(10^{-2})$ in the next 5 billion years (Gyr). \citet{laskar2009existence} and \citet{zeebe2015highly} each performed $\mathcal{O}(10^3)$ N-body simulations that roughly confirmed the secular results. Specifically, we can use the results from \citet{laskar2009existence} to calculate the probability of a Mercury instability event in the next 5 Gyr as $8.0\pm 1.8 \times 10^{-3}$ and those from \citet{zeebe2015highly} to calculate it as $6.3\pm 2.0 \times 10^{-3}$, assuming 1-$\sigma$ errors calculated using Eq.~(\ref{eq:sigma_p}).

Since a Mercury instability event would represent a major milestone in Solar System evolution, it is interesting to consider the probability of such events in comparison to other milestones. \citet{laskar1994large} chose a 5 Gyr time frame because that is the remaining main-sequence lifetime of the Sun. Another key milestone will occur in roughly 2 Gyr when Earth loses its habitability either through water loss or a runaway greenhouse \citep{wolf2015evolution}, which motivates trying to evaluate the probability of Mercury instability events at earlier times. Although the simulations of \citet{laskar2009existence} and \citet{zeebe2015highly} allow us to calculate the probability of Mercury instability events in the next 5 Gyr, these events are too rare in the next 3 Gyr for their probability to be estimated accurately from their simulation suites. Specifically, in the first 3 Gyr there are only 2 occurrences out of 2501 simulations performed by \citet{laskar2009existence} and only 1 occurrence out of 1600 simulations performed by \citet{zeebe2015highly}. It would be possible to improve event statistics by greatly increasing the number of simulations performed, but the associated computational cost makes this an undesirable solution.

In this paper we will use a rare event sampling technique to obtain accurate estimates of the probability that Mercury's orbit becomes unstable in the next 2 to 3 Gyr without increasing the computational cost. This method is called Quantile Diffusion Monte Carlo \citep[QDMC,][]{webber2019practical}. Using QDMC we pause the simulations regularly and check a ``reaction coordinate'' that indicates progress toward the desired event. We then rank the simulations in terms of this reaction coordinate. Next we preferentially restart multiple copies of simulations whose rank progressed more toward the desired event with slightly different initial conditions (splitting) and preferentially stop simulations whose rank progressed less toward the desired event (killing). We both split and kill probabilistically, and throughout this process we carefully track the splitting and killing history so that we can recover the statistics of the underlying model. As we do this, we maintain a constant total number of simulations and therefore constant computational cost. The net effect of QDMC is that we devote a vastly disproportionate amount of computational power to simulating the rare event of interest, and as a result greatly improve our estimate of its probability and the uncertainty in this probability. Using QDMC, we are able to obtain unbiased estimates of the probability of a Mercury instability event between 2~Gyr and 3~Gyr in the future at a computational cost that is up to 100 times less than direct numerical simulation. 

Our work builds on two previous efforts in this area that have also used splitting and killing \citep{laskar1994large,batygin2008dynamical}. Critically, both of those efforts merely split the simulation that had progressed the most and killed the others, which prevented them from reconstructing event statistics in the underlying model. Moreover, the underlying model of \citet{laskar1994large} was secular and the model of \citet{batygin2008dynamical} did not include general relativity, which has a dramatic effect on Mercury instability event statistics \citep{laskar2008chaotic,laskar2009existence}. In this paper we will use the Rebound N-body code \citep{rein2012rebound} including expansion terms that represent general relativity \citep{Tamayo2020reboundx}. Finally, both \citet{laskar1994large} and \citet{batygin2008dynamical} used Mercury's eccentricity to measure progress toward Mercury instability events (as a reaction coordinate). We will show that Mercury's instantaneous or time-averaged eccentricity is a poor predictor of Mercury instability events $\sim$0.5~Gyr in the future. Instead, we use the range (range = max - min) of the Mercury-Venus Minimum Orbit Intersection Distance (MOID) 0.4-0.8~Gyr before a potential instability event as our reaction coordinate. The Mercury-Venus MOID range can be calculated quickly and is correlated with the Mercury-Venus orbital energy. We find that large variations in MOID are good predictors of future instability in Mercury's orbit, possibly because they are a sign of energy being pumped into and out of Mercury's orbit by other planets. 

This paper is organized as follows. We describe our methods in Section~\ref{sec:methods}, including the REBOUND N-body code that we will use (Section~\ref{sec:rebound}), our direct numerical simulations (Section~\ref{sec:dns}), our implementation of QDMC (Section~\ref{sec:qdmc}), and our choice of reaction coordinate (Section~\ref{sec:methods:reaction_coordinate}). We give our results in Section~\ref{sec:results}, including showing that our direct numerical simulation results are consistent with previous work (Section~\ref{sec:dns-results}) and demonstrating the improvement in Mercury instability event statistics we can obtain with QDMC (Section~\ref{sec:qdmc-results}). We discuss the limitations and implications of our work in Section~\ref{sec:discussion} and conclude in Section~\ref{sec:conclusion}.

\section{Methods} \label{sec:methods}

Our implementation of QDMC for REBOUND as well as the data and code to produce all of the figures in this paper are permanently archived in the Knowledge@UChicago repository at the URL  https://knowledge.uchicago.edu/record/3323?\&ln=en.

\subsection{REBOUND}\label{sec:rebound}

We perform all simulations in this paper using the REBOUND N-body code's \citep{rein2012rebound} \texttt{WHCKL} integration scheme \citep{Rein2019WHCKL}, which applies symplectic correctors \citep{Wisdom2006} and a modified kick step \citep{WHT1996} to a classic Wisdom-Holman scheme \citep{WH1991} in order to achieve an error scaling of ${\cal O}(\epsilon dt^{18} +\epsilon^2dt^{4} +\epsilon^3dt^{3})$ where $\epsilon\sim10^{-3}$ in the case of the solar system. We adopt a fixed time step of  $dt=\sqrt{65}\approx8.062$ days. This is small enough to ensure that Mercury's orbit is properly resolved and irrational in order to avoid step-size resonances. Our integrator configuration matches the configuration used for the integrations described in \citet{brown2020repository}. Our integrations contain all 8 Solar System planets and include an approximation of general relativity \citep{Tamayo2020reboundx}. We start the simulations from current Solar System conditions with a Gaussian random perturbation to Mercury's x-coordinate position with a length scale of 1~cm. We stop the simulations if Mercury and Venus come within 0.01 AU of each other (the length scale of Venus's Hill sphere), which we use to define a Mercury instability event. This is a slightly more inclusive definition of a Mercury instability event than those used by \citet{laskar2009existence} and \citet{zeebe2015highly}, since they traced Mercury until a collision occurred. 

As described by \citet{zeebe2015dynamic}, there are many challenging numerical issues associated with the performance of N-body integrators. These issues become especially apparent when planets' eccentricities become high or when they approach each other. For example, the symplectic method can become unstable and may introduce artificial chaos unless the time step is small enough
to always resolve perihelion \citep{zeebe2015dynamic,wisdom2015resolving}. The REBOUND configuration we are using should be accurate until shortly before a Mercury instability event occurs, which is consistent with the fact that our direct numerical simulation estimates of the probability of Mercury instability events are similar to those of \citet{laskar2009existence} and \citet{zeebe2015highly} (Section~\ref{sec:dns-results}). In any case, what is most important for this paper is that QDMC can produce unbiased results at reduced computational cost compared to direct numerical simulation with the same underlying model. 

\subsection{Direct Numerical Simulation}\label{sec:dns}

We perform 1008 direct numerical simulations of the Solar System for 5 Gyr. In order to make the underlying model of the direct numerical simulations exactly the same as the QDMC simulations described in Section~\ref{sec:qdmc}, we apply a Gaussian random perturbation to Mercury's position with a length scale of 1~cm every 0.2~Gyr in our direct numerical simulations. Given the chaotic nature of Solar System dynamics \citep{laskar1989numerical}, we do not expect these perturbations to change the statistics of Mercury instability events. They may, however, affect which particular simulations experience a Mercury instability event.

Our different Mercury simulations evolve independently, so the number of observed instability events is a binomial random variable. 
Let $p_e$ denote the true probability of instability,  $N$ the number of simulations, and $N_e$ the number of observed instability events.
We estimate $p_e$ using the maximum likelihood formula
$\hat{p}_e = N_e/N$, where hat denotes an estimated quantity,
and estimate 1-$\sigma$ error bars using
\begin{equation}
\hat{\sigma}_p = \sqrt{\frac{\hat{p}_e (1 - \hat{p}_e)}{N}}.
\label{eq:sigma_p}
\end{equation}
We will use the estimates $\hat{p}_e$ and $\hat{\sigma}_p$ for our direct numerical simulation results, as well as those of \citet{laskar2009existence} and \citet{zeebe2015highly}.

\subsection{QDMC}\label{sec:qdmc}

\emph{Background}.
The basic principle underlying QDMC is that an ensemble of simultaneous model simulations can be pushed toward a rare event of interest by occasional duplication of some ensemble members and removal of others (resampling).  Incarnations of this ``splitting'' idea appear in print at least as early as the 1950's \citep{KahnHarris:1951:splitting,RosenbluthRosenbluth:1955:SIS}. Since then, many splitting schemes have been used in rare event simulation problems in physics \citep{Grassberger:1997:splitting}, chemistry \citep{HUBER1996we,AllenFrenkel:2006:FFS,Warmflash:2007:NEUS,GuttenbergDinner:2012:STPS}, computer science \citep{VillenAltamiranoVillenAltamirano:1991:RESTART,HarasztiTownsend:1999:DPR}, statistics \citep{DelMoral2005rare,CerouGuyader:2007:AMS} and geophysics \citep{Ragone2018heatwaves, webber2019practical,ragone2021rare}.  

The first and most important step in the application of a rare event simulation technique is the selection of a reaction coordinate.  The reaction coordinate quantifies the progress of an ensemble member toward the rare event, and the change in its value since the last resampling determines the likelihood that the ensemble member will be duplicated or removed.  In most situations the optimal reaction coordinate for computing the probability of a rare event is the probability of the event as a function of the current state of the system \citep[e.g.,][]{DupuisWang:2004:Stochastics,DeanDupuis:2009:splitting,DeanDupuis:2011:AnnOpRes,VandenEijndenWeare:2012:RareEvent,webber2020splitting}. However, in many cases a very rough descriptor of this probability is sufficient.  In applications of real scientific interest, determination of an effective reaction coordinate often requires a non-trivial computational interrogation of the system \citep{Antoszewski2020insulin}.

Quantile DMC \citep[QDMC,][]{webber2019practical} is a recently proposed splitting method that builds on previous approaches by improving the robustness properties.
As opposed to other splitting and killing schemes  \citep[e.g.,][]{giardina2006direct,Ragone2018heatwaves},
QDMC is based on the sorting of trajectories along a reaction coordinate.
First, QDMC sorts simulations from the lowest reaction coordinate value to the highest reaction coordinate value.
Next, QDMC randomly splits simulations that have moved from lower to higher relative positions in the reaction coordinate
and randomly kills simulations that have moved from higher to lower relative positions in the reaction coordinate.
Because QDMC uses relative positions along a reaction coordinate instead of absolute positions,
it is more robust to the choice of reaction coordinate \citep{webber2019practical}.

\emph{Algorithmic details}.
QDMC starts with an initialization step and then iterates over a quantile transformation step, a splitting and killing step, and a forward evolution step, as detailed below.
\begin{enumerate}
\item \textbf{Initialization.} Generate independent simulations $X_{t_0}^1, \ldots, X_{t_0}^N$ and assign equal weights to each simulation, i.e., $w_{t_0}^i = 1/N$ for each $i = 1, \ldots, N$.
\item \textbf{Iteration.} Apply steps (a)-(c) at uniformly spaced times $t = t_0, t_0 + \Delta, t_0 + 2\Delta, \ldots$. 
\begin{enumerate}
    \item \textbf{Quantile transformation.} Evaluate the reaction coordinate $\theta\left(X_t^i\right)$ for each simulation $X_t^1, X_t^2, \ldots$,
    and find a permutation $\alpha\left(1\right), \alpha\left(2\right), \ldots$ that reorders the reaction coordinates from lowest to highest, i.e.,
    \begin{equation}
    \theta\left(X_t^{\alpha\left(1\right)}\right) \leq \theta\left(X_t^{\alpha\left(2\right)}\right) \leq \cdots.
    \end{equation}
    Next, for each simulation $X_t^1, X_t^2, \ldots$, define a rescaled reaction coordinate
    \begin{equation}
        \widehat{\theta}_t\left(X_t^i\right) = \Phi^{-1}\left( \sum_{j\colon\theta\left(X_t^j\right) < \theta\left(X_t^i\right)} w_t^j + \frac{w_t^i}{2}\right),
    \end{equation}
    where $\Phi^{-1}$ is the inverse of the Gaussian cumulative distribution function $\Phi\left(x\right) = \int_{-\infty}^x \left(2\pi\right)^{-1/2} \exp\left(-z^2/2\right) \mathop{dz}$.
    This is done to ensure that the rescaled reaction coordinates are approximately Gaussian distributed.
 
    \item \textbf{Splitting and killing}. 
    Divide the simulations into stable and unstable trajectories.
    Set aside the unstable trajectories for statistical analysis later:
    these trajectories remain part of the statistical sample but their forward evolution is stopped.
    Split each stable simulation $X_t^i$ into $N_t^i$ identical replicas, where the numbers $N_t^i$ are random chosen to satisfy
    \begin{equation}
        \mathbb{E} N_t^i = \frac{N w_t^i \exp\left(k_t \widehat{\theta}_t\left(X_t^i\right)\right) }{ \sum_{j\colon X_t^j \text{ is stable}} w_t^j \exp\left(k_t \widehat{\theta}_t\left(X_t^j\right)\right) }
        \quad \text{and} \quad
        \sum_{i\colon X_t^i \text{ is stable}} N_t^i = N,
    \end{equation}
    for an intensity parameter $k_t > 0$.
    Assign the children of $X_t^i$ weights summing to $w_t^i$ in expectation,
    following the procedure in Appendix \ref{app:implementation}.
    \item \textbf{Forward evolution}. 
    Advance the stable trajectories forward until the next resampling time, using independent random seeds.
\end{enumerate}
\item \textbf{Statistical estimation}. At any time, use the weighted measure $\sum_i w_t^i \delta_{X_t^i}$ to produce unbiased estimates with respect to the simulation model, e.g.,
\begin{equation}
    \textup{Prob}\left\{\textup{instability at time $t$}\right\} \approx 
    \sum_{i\colon X_t^i \text{ is unstable}} w_t^i.
\end{equation}
The sum of the weights $\sum_i w_t^i$ is exactly one,
and these weights provided the estimated probabilities of different events occurring.
\end{enumerate}

In QDMC, the dynamics are determined by a rescaled reaction coordinate $\widehat{\theta}$
and by an
intensity function $k_t > 0$ (in standard deviation units), which controls the strength of the splitting and killing.
Using direct numerical simulations,
the distribution of $\widehat{\theta}_t\left(X_t^1\right), \widehat{\theta}_t\left(X_t^2\right), \ldots$ would be nearly $\mathcal{N}\left(0, 1\right)$.
However, QDMC drives
the distribution of rescaled reaction coordinates $\widehat{\theta}_t\left(X_t^1\right), \widehat{\theta}_t\left(X_t^2\right), \ldots$ to nearly $\mathcal{N}\left(k_t, 1\right)$.
In this sense, QDMC pushes the distribution of reaction coordinate values $k_t$ standard deviation units higher, as compared to the underlying simulation model.

\emph{Parameter choices}.
In typical QDMC applications, we gradually increase the intensity function $k_t$ leading up to a target time $t = t_{\text{target}}$.
To obtain an explicit formula for $k_t$, 
we model the rescaled reaction coordinate $\hat{\theta}_t$ as a Gaussian linear process
\begin{equation}
\label{eq:assumption}
\begin{cases}
    & \mathop{d\hat{\theta}_t} = -\left(\hat{\theta}_t \slash \tau\right) \mathop{dt} + \sqrt{2 \slash \tau} \mathop{dW}, \quad t > 0 \\
    & X_0 = 0,
\end{cases}
\end{equation}
with a decorrelation timescale $\tau$,
and we optimize $k_t$
to yield the minimal variance estimate for
$\mathbb{E} \exp\left(k_{\max} X_{t_{\text{target}}}\right)$ 
\citep[see e.g.][]{CerouGuyader:2007:AMS,webber2020splitting}.
The resulting intensity function is
\begin{equation}
    k_t = k_{\max} \exp\left(\frac{t - t_{\text{target}}}{\tau} \sqrt{1 - \exp\left(\frac{-2t}{\tau}\right)}\right).
\end{equation}
The tuning parameters for QDMC are thus the reaction coordinate $k_t$, the
decorrelation timescale $\tau$, 
the intensity threshold $k_{\max}$,
and the resampling times leading up to $t_{\textup{target}}$.

QDMC produces unbiased, convergent estimates regardless of the specific parameters,
yet the choice of parameters does affect QDMC's efficiency at converging to the correct statistics.
The main parameter determining QDMC's efficiency is the reaction coordinate $\theta_t$,
which is discussed
at length in Section \ref{sec:methods:reaction_coordinate}.
QDMC is comparatively less sensitive to the other parameters, and we follow the recommendations given in \cite{webber2019practical} as described below:

\begin{itemize}
\item The decorrelation timescale $\tau$ should be tuned based on the timescale for large random changes in the rare event probability.
Since Mercury's eccentricity can increase from $.3$ to $.8$ over a timescale of $.5$~Gyr \citep{laskar2009existence,zeebe2015highly} and since small Gaussian perturbations can potentially push Mercury toward or away from such spiraling instability,
we set $\tau = .5$~Gyr.
However, as explained in Sec. IIE of \cite{webber2019practical},
QDMC is not very sensitive to the decorrelation timescale, and over-estimating or under-estimating the timescale by a factor of two causes less than a $20\%$ increase in QDMC's error.
\item The intensity threshold $k_{\max}$ should be tuned based on the approximate magnitude of the rare event probability.
Because the simulation results of \cite{laskar2009existence} and \cite{zeebe2015highly} suggest the probability of Mercury becoming unstable $2$ -- $3$~Gyr into the future is between $10^{-4}$ and $10^{-3}$, we set the intensity threshold to be $k_{\max} = 3$, which is appropriate for estimating rare events with probabilities ranging from $3 \times 10^{-5}$ -- $2 \times 10^{-2}$ (i.e., rare excursions of $2-4$ standard deviations away from the mean).
\item There should be a modest number of resampling times ($\leq 10$)
during the interval $\left[t_{\text{target}} - 2 \tau, t_{\text{target}}\right]$ so that QDMC
achieves its maximum efficiency at the target time $t_{\text{target}}$.
Here, we use the specific resampling times
\begin{equation}
    t_{\text{target}} - 1\text{ Gyr},\quad
    t_{\text{target}} - .8\text{ Gyr},\quad
    t_{\text{target}} - .6\text{ Gyr},\quad
    t_{\text{target}} - .4\text{ Gyr},\quad
    t_{\text{target}} - .2\text{ Gyr},
\end{equation}
and we run three separate QDMC trials using 
$t_{\text{target}} = 2.4\text{~Gyr}$,
$t_{\text{target}} = 2.8\text{~Gyr}$,
and $t_{\text{target}} = 3.2\text{~Gyr}$.
This schedule is slightly different from the recommendation given in \citep{webber2020splitting},
which suggests increasing the frequency of resampling times as $t \rightarrow t_{\text{target}}$.
However, here we are limited in how short we can make the intervals between resampling times 
because it takes approximately $0.1-0.15$~Gyr for orbits with small initial perturbations to spread apart in eccentricity values (Fig.~\ref{fig:trajectories-gr-dns}).
To ensure adequate statistical variation among our samples, we apply splitting and killing at regularly spaced $0.2$~Gyr intervals.
\end{itemize}

\emph{Variance estimation}.
We provide
two variance estimators for QDMC, one which overestimates the variance and another which underestimates the variance.
For any statistical estimate of the form
$\hat{f} = \sum_i w_t^i f\left(X_t^i\right)$,
we can overestimate the variance using the following \emph{pessimistic} variance estimator
\begin{equation}
    \hat{\sigma}^2_{\text{pess}} = \sum_i \left|\sum_{\text{anc}\left(X_t^j\right) = i} w_t^j f\left(X_t^j\right)\right|^2 - \hat{f}^2.
\end{equation}
Here, $\text{anc}\left(X_t^j\right)$ denotes the index for the original ancestor of the simulation $X_t^j$ traced back to time $t = 0$.
We can underestimate the variance using the following \emph{optimistic} variance estimator
\begin{equation}
    \hat{\sigma}^2_{\text{opt}} = \sum_i \left| w_t^i f\left(X_t^i\right)\right|^2 - \hat{f}^2.
\end{equation}
To intuitively explain the difference between these two variance estimators, 
we remark that the optimistic estimator treats each individual simulation as a single data point as in importance sampling \citep{liu2008monte}, 
while the pessimistic estimator treats each family of simulations as a single data point.
However, the true sample size lies somewhere between these two extremes.
Simulations with a lot of shared ancestry are often quite correlated with one another;
however, simulations that diverged many generations ago are hardly correlated at all.
While this explanation captures the essential intuition, we note that variance estimation for splitting schemes remains an open area of mathematical investigation.
As of the present writing, the pessimistic variance estimator is known to overestimate the variance for large sample sizes \citep{webber2019practical},
and the optimistic variance estimator is found empirically to underestimate the variance \citep{webber2022variance}.
Further work is needed to close the gap between these two variance estimators.

Once we have obtained an estimate of the probability of a Mercury instability event and the associated error using QDMC, we can estimate the number of direct numerical simulations that would be necessary to produce a similarly accurate estimate using Eq.~(\ref{eq:sigma_p}) as follows
\begin{equation}
    N \approx \frac{\hat{p}_e(1-\hat{p}_e)}{\hat{\sigma}_p^2}.
    \label{eq:N}
\end{equation}
Since all of our QDMC implementations involve 1008 simulations, we can estimate the speed-up provided by QDMC as $\frac{N}{1008}$, where $N$ is given by Eq.~(\ref{eq:N}).

\emph{Improvements over past work}.
Compared to our previous application of QDMC to compute intensity statistics for tropical cyclones \citep{webber2019practical}, we note two improvements.
First, following recent research on the stability of splitting and killing schemes \citep{webber2020splitting},
we constrain the sum of the weights $\sum_i w_t^i$ to be exactly one, which promotes greater interpretability and ensures stability in the limit as $t \rightarrow \infty$.
Second, while before we only used the pessimistic variance estimator $\hat{\sigma}^2_{\text{pess}}$ to compute error bars,
here we also provide the optimistic variance estimator $\hat{\sigma}^2_{\text{opt}}$,
and we compare the two estimators.

\subsection{Choice of reaction coordinate}
\label{sec:methods:reaction_coordinate}
The ideal reaction coordinate would encompass all early warning signs that a Mercury instability event may occur.
Identifying such an ideal reaction coordinate is extremely difficult;
however, we can search for a substitute reaction coordinate that is easy to compute and
indicates the likely occurrence of a Mercury instability event.
In this work, we consider the median and range (range = max - min) of the Mercury-Venus Minimum Orbit Intersection Distance (MOID, measured in AU), the Venus-Earth MOID, and Mercury's eccentricity as potential reaction coordinates.

To assess the usefulness of potential reaction coordinates,
we analyze our direct numerical simulation data set.
This data set encompasses $996$ trajectories that remained stable throughout the entire $5$ Gyr integration and $12$  trajectories in which a Mercury instability event occurred.
We consider times $400-800$ Myr before the simulation ended either at $5$ Gyr or in a Mercury instability event.
Then, using the time series for the Mercury-Venus MOID, the Venus-Earth MOID, and Mercury's eccentricity,
we compute the median and the range
for statistical analysis.

\begin{figure}[ht!]
\centering
\includegraphics[width=0.5\linewidth]{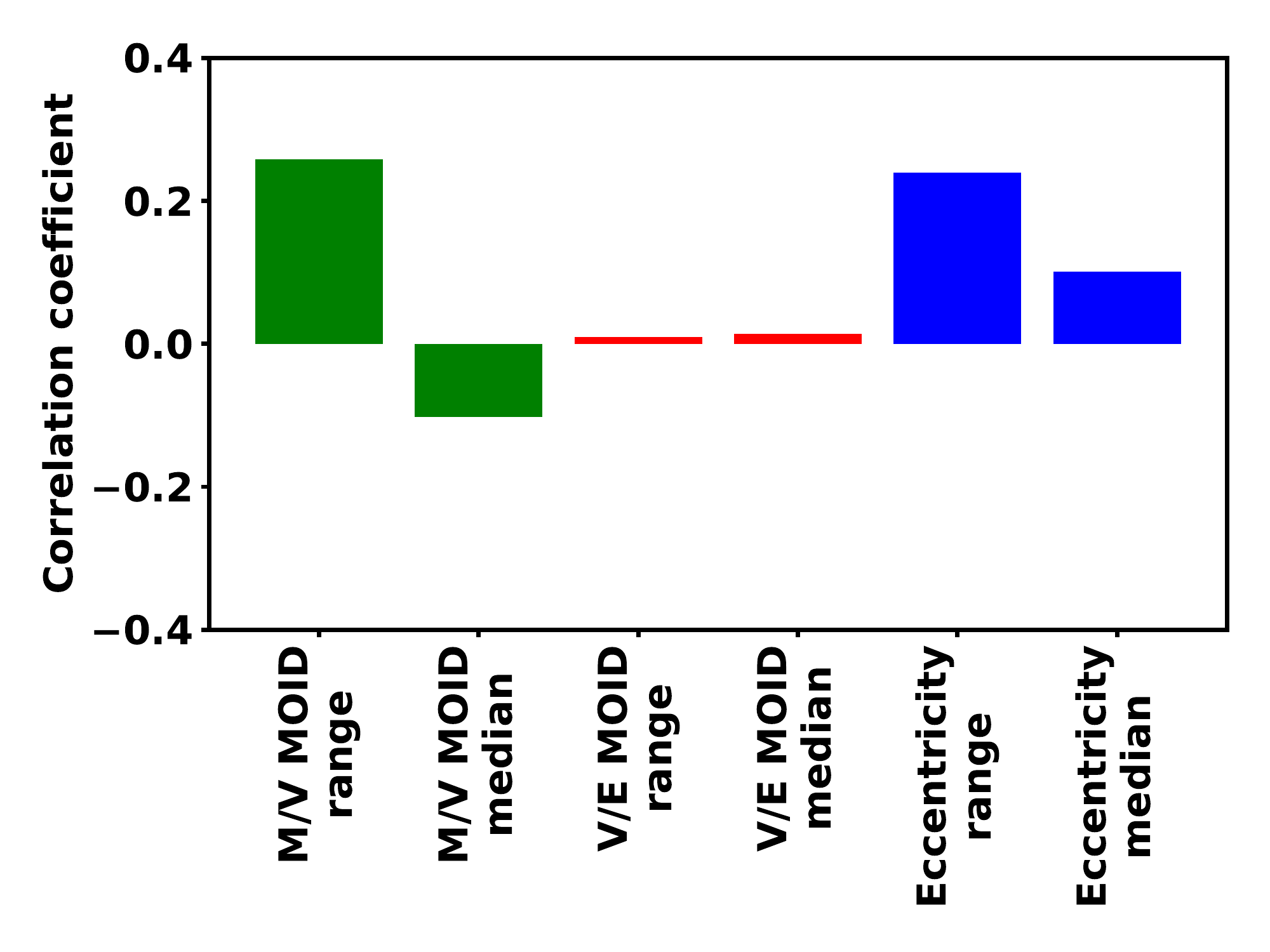}
\caption{\textbf{Mercury-Venus MOID is the most effective predictor of Mercury instability events $400-800$ Myr in the future.}
Barplot shows correlation coefficients between Mercury stability/instability and $6$ potential reaction coordinates, calculated $400-800$ Myr before the simulation end. The potential reaction coordinates are the median and range of the Mercury-Venus MOID, the Venus-Earth MOID, and Mercury's eccentricity.
\label{fig:predicting}}
\end{figure}

As a first statistical test,
we calculate correlation coefficients
between the binary variable indicating stability/instability and the $6$ predictor variables,
as shown in Figure \ref{fig:predicting}.
This comparison suggests that the Mercury-Venus MOID range and Mercury's eccentricity range
are the strongest instability predictors,
with correlation coefficients of $r = .26$
and $r = .24$ respectively.
The medians of Mercury-Venus MOID and Mercury eccentricity are comparatively poor predictors of instability ($r = -.10$ and $r = .10$),
while the median and range of Venus-Earth MOID have nearly no predictive potential ($r = .01$ and $r = .01$). Note that this implies that a simulation with large variation in its Mercury-Venus MOID is typically more likely to experience a Mecury instability event $400-800$ Myr in the future than a simulation that has low Mercury-Venus MOID without much variation.

As a second statistical test, we run a Lasso logistic regression \citep{tibshirani1996regression,scikit-learn}
to predict future instability as a sparse linear combination of the $6$ predictor variables.
With a regularization parameter of $C = 0.01$, the resulting model is
\begin{equation}
\label{eq:model}
    \log\left(\frac{p}{1-p}\right)
    = -.224 + .569\, X_{\textup{MOID}} + .159 \,X_{\textup{eccen}},
\end{equation}
where $p$ is the probability of instability, while $X_{\textup{MOID}}$ and $X_{\textup{eccen}}$ represent the Mercury-Venus MOID range and Mercury's eccentricity range, both normalized to have a mean of zero and a standard deviation of one.
This model suggests that Mercury-Venus MOID range
heavily influences the probability of instability,
while Mercury's eccentricity range makes a smaller secondary contribution.

As a result of this analysis, we use the Mercury-Venus MOID range calculated over $400$ Myr of data as our reaction coordinate. Although we could potentially obtain better results by combining MOID data and eccentricity data into a single instability predictor,
Equation \eqref{eq:model} suggests that the role played by eccentricity in the optimal predictor is likely to be comparatively small.
Using MOID data alone also leads to greater interpretability.

We do not have a complete explanation for why the MOID range is an effective predictor of future Mercury instability events. Here we simply note that the Mercury-Venus MOID is strongly correlated with the Mercury-Venus orbit-averaged interaction energy. Given this feature of the MOID and the fact that Mercury's eccentricity excitation is known to result from secular (i.e., orbit-averaged) dynamical interactions in which Venus plays a prominent role \citep[e.g.,][]{LithwickWu2011secular_chaos}, it is reasonable that the MOID range correlates with instability probability. The mechanism may be that large variations in the Mercury-Venus orbit-averaged interaction energy, captured by large variations in the Mercury-Venus MOID, are a sign of energy being pumped into and out of the system in a way that can lead to future instability.

\section{Results} \label{sec:results}

\subsection{Direct Numerical Simulation}\label{sec:dns-results}

\begin{figure}[ht!]
\centering
\includegraphics[width=0.5\linewidth]{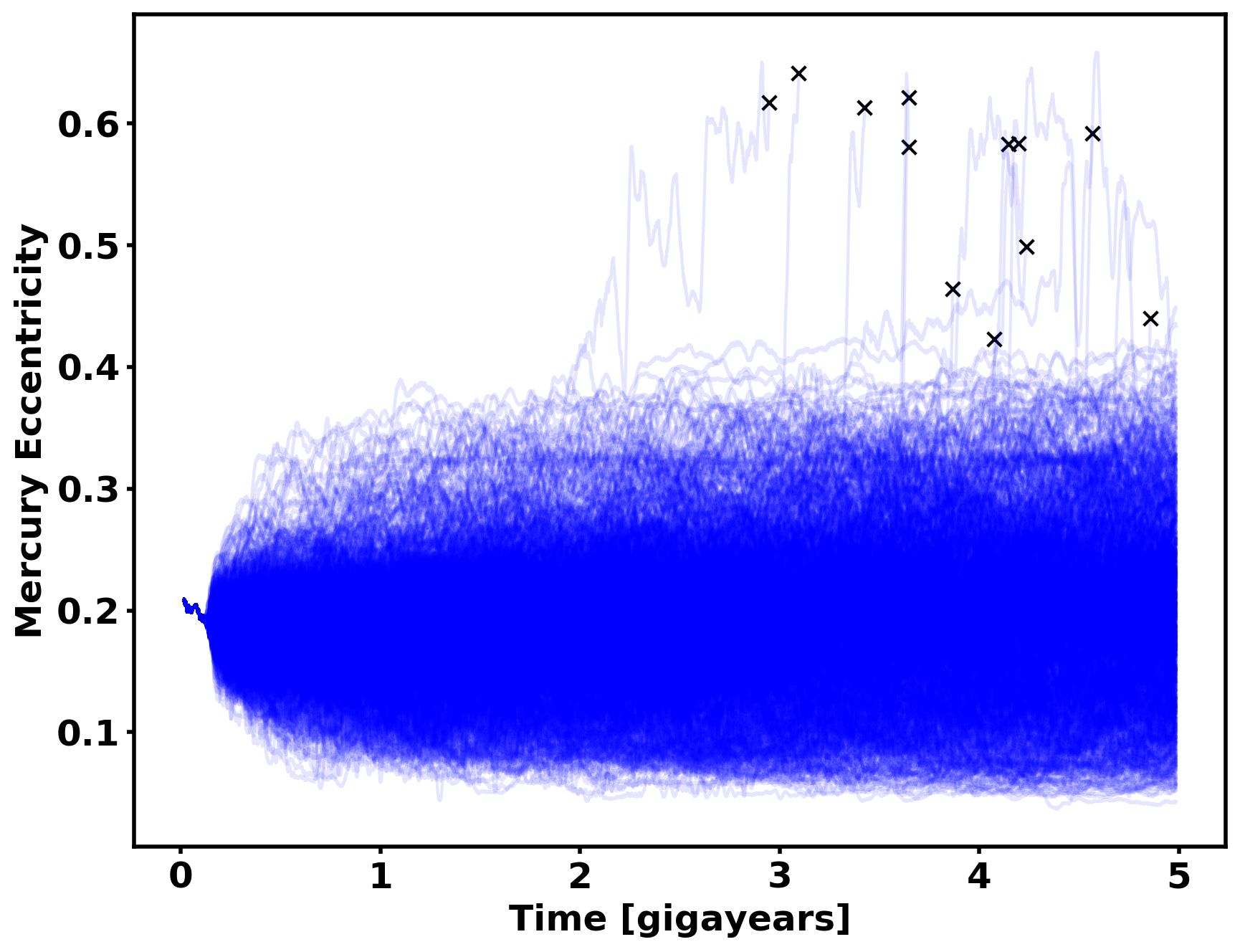}
\caption{\textbf{Mercury's eccentricity from 1008 Direct Numerical Simulation (DNS) 5 Gyr integrations of the solar system.} The eccentricity plotted is a 30-Myr running-average. We add perturbations to Mercury's position on the scale of 1 cm every 200 Myr in these integrations for similarity with our Quantile Diffusion Monte Carlo (QDMC) simulations, but this should not affect event statistics. The x's signify a close encounter between Mercury and Venus, at which point we stop our simulations. The x's occur in this plot when Mercury's eccentricity is relatively small because of the running average and our output frequency.}
\label{fig:trajectories-gr-dns}
\end{figure}

The paths of Mercury's eccentricity for the next 5 Gyr in our 1008 direct numerical simulations are shown in Fig.~\ref{fig:trajectories-gr-dns}. We apply a 30-Myr running average to Mercury's eccentricity in this plot so that individual simulations can be discerned. All of the simulations stay very near to each other for the first $\sim$0.1~Gyr, at which point they start to diverge. For the most part, the spread in Mercury's eccentricity increases slowly and roughly diffusively \citep{laskar2008chaotic}. Some simulations jump out of this slow spread in Mercury's eccentricity, and eventually result in a close encounter between Mercury and Venus. Most of these Mercury instability events happen suddenly, with very little advance warning from Mercury's eccentricity. Since a reaction coordinate must predict a rare event long in advance of the event occurring, this explains why Mercury's eccentricity (rather than MOID range or eccentricity range) is not a good reaction coordinate for Mercury instability events.

\begin{figure}[ht!]
\centering
\includegraphics[width=0.5\linewidth]{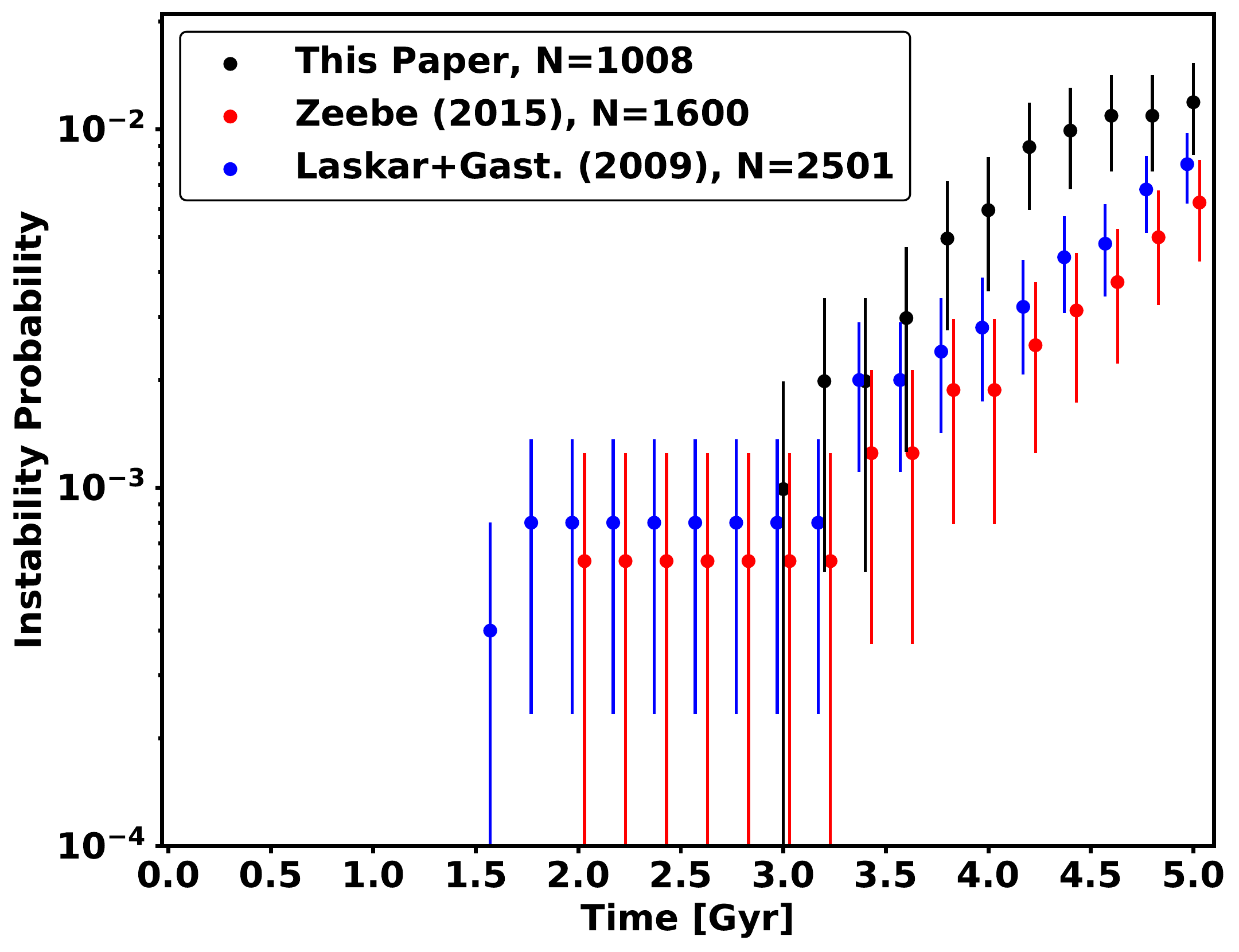}
\caption{\textbf{Our Direct Numerical Simulation (DNS) results are similar to previous studies, but slightly higher.} An estimate of the probability that Mercury has a close encounter with Venus is plotted as a function of time for our DNS simulations (black), the DNS simulations of \citet{zeebe2015highly} (red), and the DNS simulations of \citet{laskar2009existence} (blue). The error plotted is the 1-$\sigma$ error calculated using Eq.~(\ref{eq:sigma_p}).}
\label{fig:dns-instability}
\end{figure}

The probability of Mercury instability events as a function of time in our direct numerical simulations is broadly similar to that of \citet{laskar2009existence} and \citet{zeebe2015highly} (Fig.~\ref{fig:dns-instability}). For all three simulation suites the probability is lower than $10^{-3}$ before 3~Gyr in the future, then grows slowly to $\mathcal{O}(10^{-2})$ at 5~Gyr in the future. The probabilities we estimate are slightly higher than those of 
\citet{laskar2009existence} and \citet{zeebe2015highly}, especially between 3.5 and 4.5 Gyr in the future. This could be due to our more generous definition of Mercury instability events (section~\ref{sec:rebound}), the fact that we do not decrease the time step at high eccentricity, and/or the chaotic nature of the system. This issue does not affect the primary goal of this project, which is to demonstrate that QDMC provides a computationally efficient unbiased estimate of Mercury instability statistics within a given model.

\subsection{QDMC}\label{sec:qdmc-results}

\begin{figure}[ht!]
\centering
\includegraphics[width=0.5\linewidth]{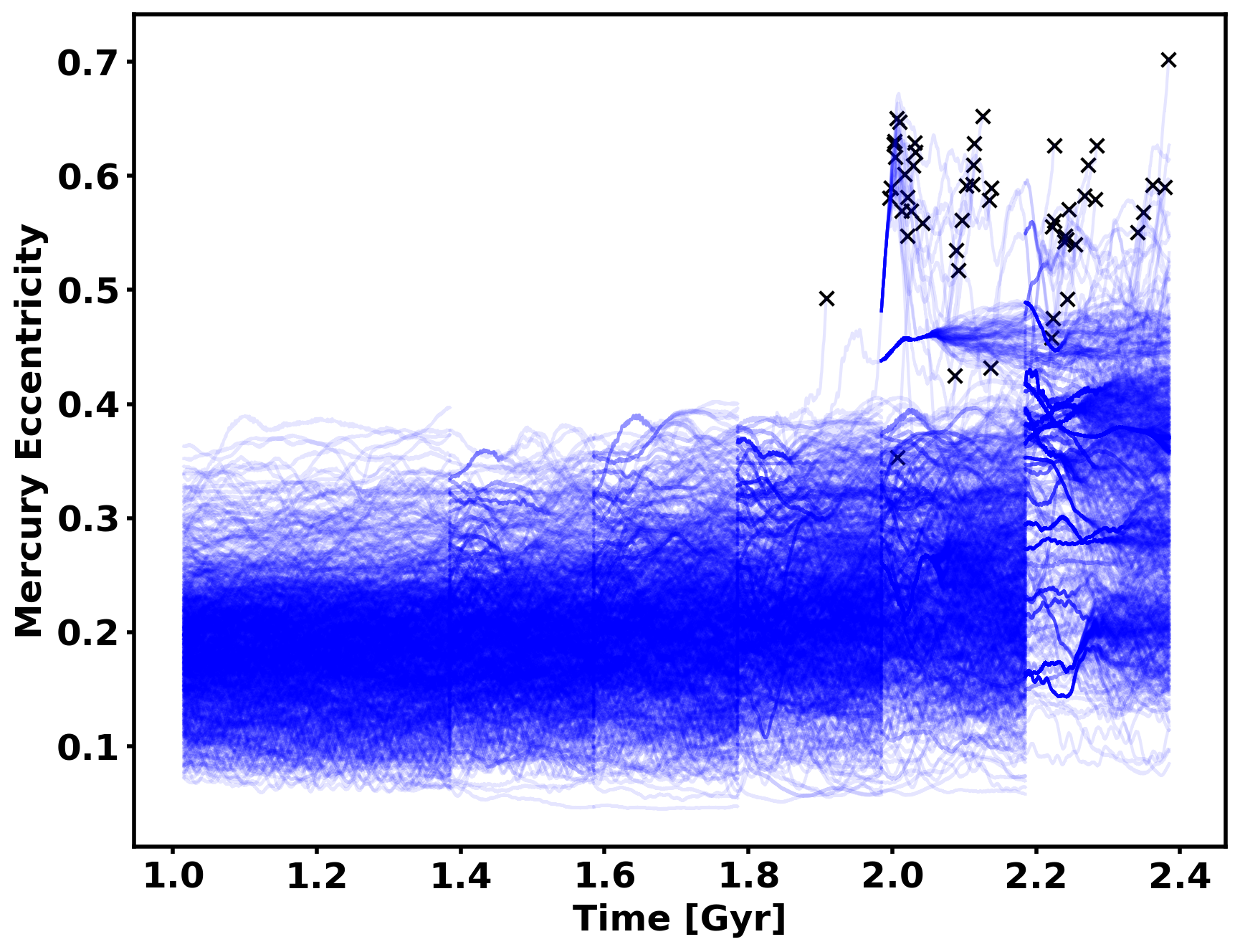}
\caption{\textbf{Quantile Diffusion Monte Carlo (QDMC) produces many more close encounters between Mercury and Venus than Direct Numerical Simulation (DNS).} Here we show our results for QDMC where splitting occurs every 0.2 Gyr, starts at 1.4 Gyr, and ends at a target time of 2.4 Gyr. The vertical axis is Mercury's eccentricity and the 30-Myr running-average of the simulations is plotted. The x's signify a close encounter between Mercury and Venus.}
\label{fig:qdmc-trajectories}
\end{figure}

QDMC produces many Mercury instability events at times before direct numerical simulation has produced any (Fig.~\ref{fig:qdmc-trajectories}). In Fig.~\ref{fig:qdmc-trajectories} we can see the effects of splitting and killing, particularly as the simulation progresses and we force the model more strongly toward Mercury instability events. When viewing Fig.~\ref{fig:qdmc-trajectories}, it is important to remember that we are not using Mercury's eccentricity as our reaction coordinate. This is why extensive splitting can occur at relatively low values of Mercury's eccentricity.

\begin{figure}[ht!]
\centering
\includegraphics[width=0.5\linewidth]{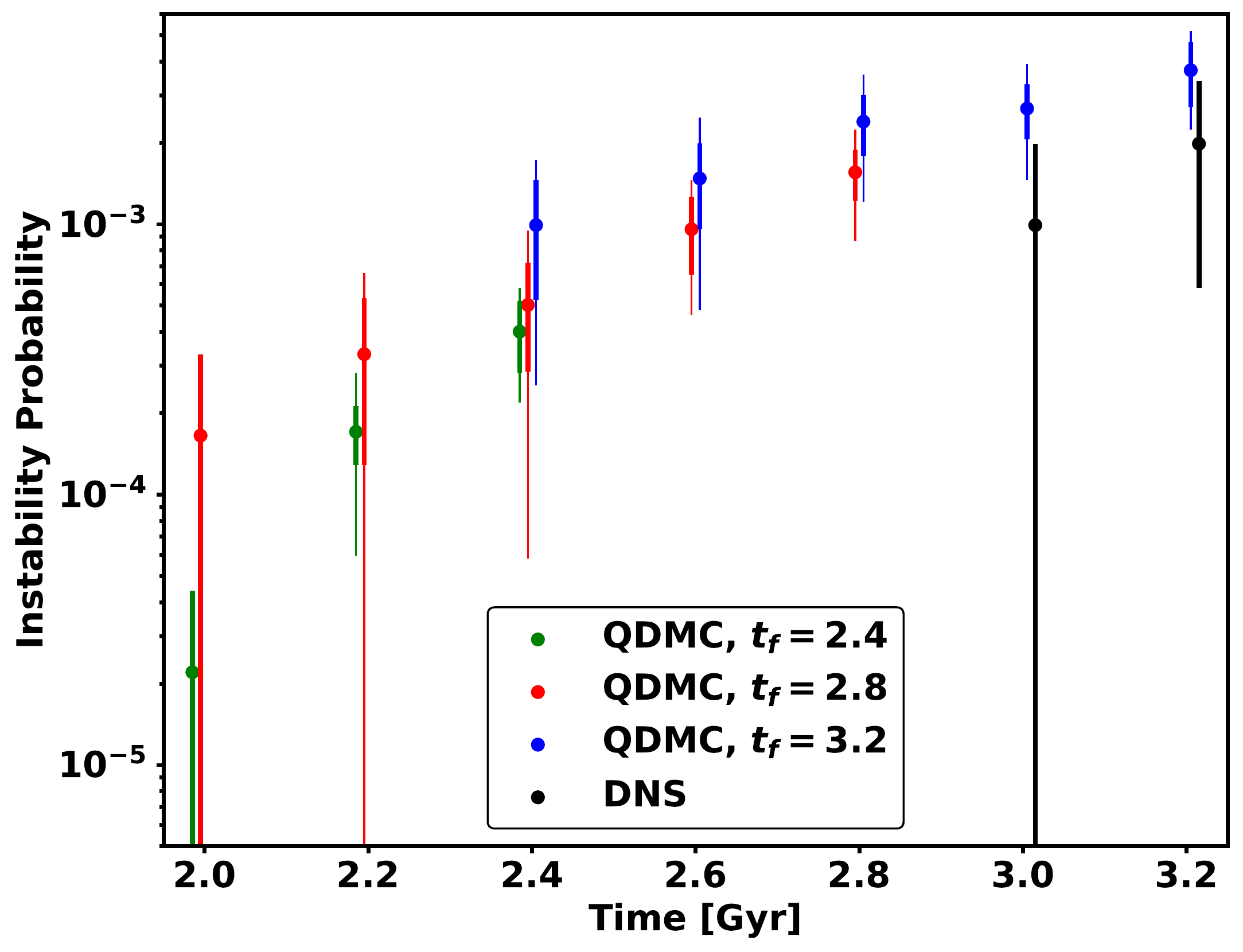}
\caption{\textbf{Quantile Diffusion Monte Carlo (QDMC) produces unbiased probability estimates.} Here we are plotting the probability that Mercury has a close encounter with Venus as a function of time for our direct numerical simulations (DNS, black), our QDMC simulations with a target time of 2.4 Gyr (green), our QDMC simulations with a target time of 2.8 Gyr (red), and our QDMC simulations with a target time of 3.2 Gyr (blue). The 1-$\sigma$ error for each data point is also plotted. We calculated the error using  Eq.~(\ref{eq:sigma_p}) for DNS. We provide an optimistic (thick) and pessimistic (thin) error estimate for QDMC, as described in section~\ref{sec:qdmc}.}
\label{fig:qdmc-instability}
\end{figure}

We plot our QDMC estimates of the Mercury instability event probability as a function of time in Fig.~\ref{fig:qdmc-instability}. 
We can compare the QDMC results against direct numerical simulation estimates at 3.0 and 3.2 Gyr in the future, 
and the results are consistent within the standard error bars.
Moreover, at times when we have QDMC implementations with different target times, the results are consistent within the error bars as well.
At 2.4~Gyr all three of our QDMC implementations provide overlapping probability estimates. 
All of this is consistent with QDMC providing an unbiased estimate of the underlying Mercury instability probability of the model. 

\begin{figure}[ht!]
\centering
\includegraphics[width=0.5\linewidth]{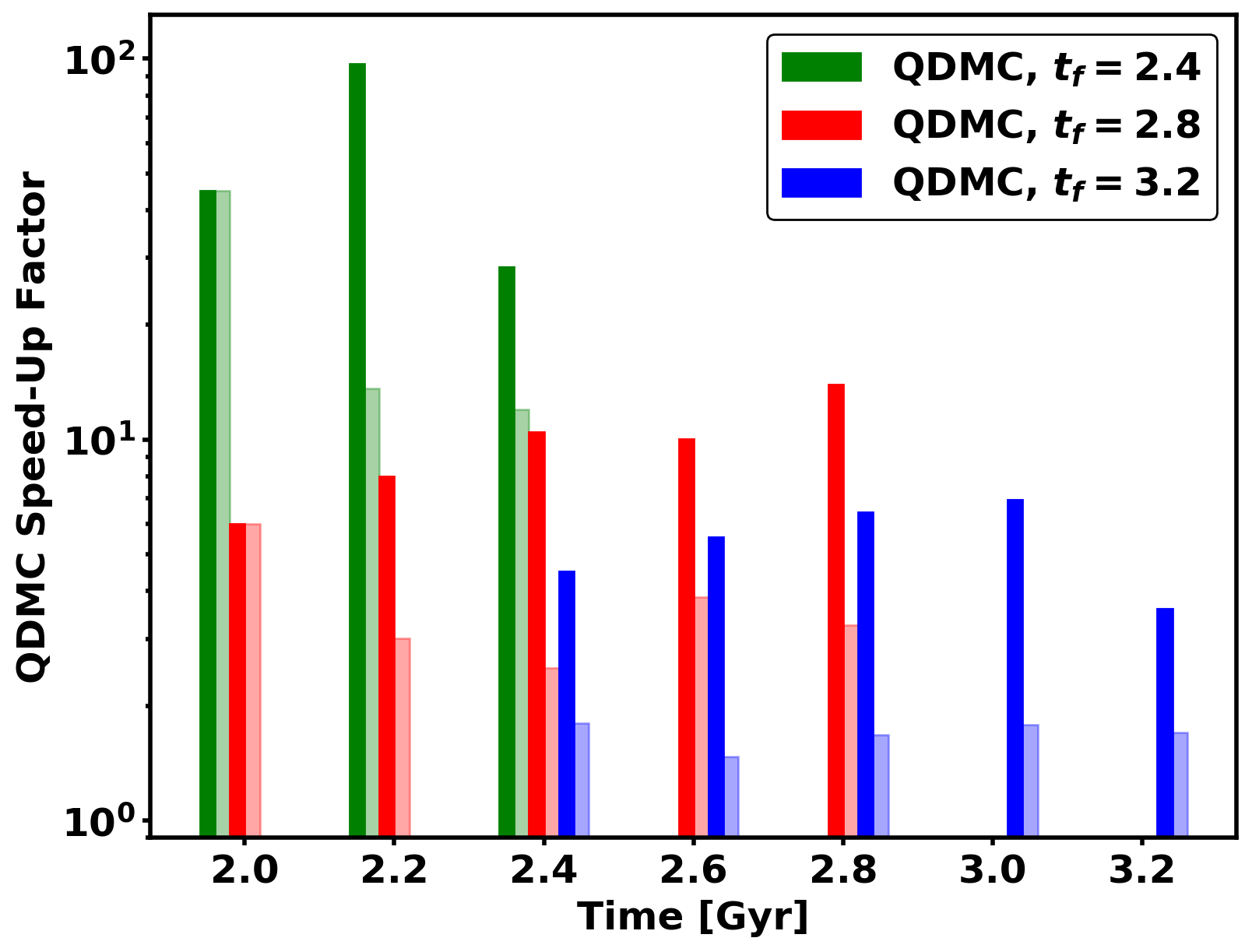}
\caption{\textbf{Quantile Diffusion Monte Carlo (QDMC) is up to 100 times faster than direct numerical simulation.} This plot shows the ratio of the inferred number of direct numerical simulations needed for the precision of QDMC (Eq.~(\ref{eq:N})) to the number used in our implementation of QDMC (1008). This tells us the computational speed-up of QDMC. We calculate speed-up for our QDMC simulations with a target time of 2.4 Gyr (green), our QDMC simulations with a target time of 2.8 Gyr (red), and our QDMC simulations with a target time of 3.2 Gyr (blue). 
We plot the speed-up for both optimistic (dark) and pessimistic (light) QDMC error estimates (section~\ref{sec:qdmc}). 
The speed-up is larger for earlier times (when the event is rarer) and reaches a value of about 100 for optimistic error bars.}
\label{fig:qdmc-speed-up}
\end{figure}

Using Eq.~(\ref{eq:N}) we can estimate the computational speed-up that QDMC provides relative to direct numerical simulation (Fig.~\ref{fig:qdmc-speed-up}). 
The speed-up is higher around 2~Gyr in the future than around 3~Gyr, which makes sense because QDMC should be more beneficial the rarer the event, and Mercury instability events are rarer at earlier times. Near 2~Gyr in the future, the speed-up reaches a maximum of about 100, assuming optimistic error bars. Even assuming pessimistic error bars, the speed-up is always greater than 10 for a QDMC target time of 2.4~Gyr. At QDMC target times of 2.8~Gyr and 3.2~Gyr the speed-up becomes progressively smaller, although it is still $\mathcal{O}(10)$ for optimistic error bars. Even assuming pessimistic error bars, QDMC always provides at least some speed-up relative to direct numerical simulation.

\section{Discussion} \label{sec:discussion}

Our direct numerical simulations as well as those of \citet{zeebe2015highly} and \citet{laskar2009existence} suggest 
that the probability of Mercury's orbit becoming unstable in the next 3~Gyr is $\mathcal{O}(10^{-3})$ and in the next 5 Gyr is $\mathcal{O}(10^{-2})$ (Fig.~\ref{fig:dns-instability}). Our QDMC results indicate that the probability of Mercury's orbit becoming unstable in the next 2~Gyr is $\mathcal{O}(10^{-4})$ (Fig.~\ref{fig:qdmc-instability}). We can use these data to speculate that the rate of increase of the log of the probability of Mercury's orbit becoming unstable decreases with time. This would suggest that the probability of Mercury's orbit becoming unstable in the next 1~Gyr is less than $\mathcal{O}(10^{-5})$. This speculation could be checked with QDMC simulations with a target time of 1~Gyr in future work, although it is possible this would require some modification of the reaction coordinate and/or more simulations than we performed here.

We performed a direct numerical simulation suite before applying QDMC (section~\ref{sec:dns}). This was necessary in order to confirm that our modelling choices yielded results consistent with previous efforts (section~\ref{sec:dns-results}) and to confirm that the QDMC results were consistent with the direct numerical simulation results where they overlapped (section~\ref{sec:qdmc-results}). Additionally, the direct numerical simulation suite was useful for developing a good reaction coordinate (section~\ref{sec:methods:reaction_coordinate}) since this was the first effort to apply QDMC to a planetary dynamics problem. It is important to note, however, that it is not necessary to perform a direct numerical simulation suite in order to apply QDMC. If you have a good reaction coordinate in hand, you can apply QDMC from the start. Moreover, if you have a reasonable guess at a reaction coordinate, you can run QDMC using it, then use the output to find an improved reaction coordinate and repeat if necessary. 

Our estimate of the speed-up associated with QDMC may seem optimistic given that we had to devote a significant amount of time and effort to finding a good reaction coordinate. It is true that such effort is necessary when a completely new problem is approached using QDMC, but it is not necessary when applying QDMC again to a similar problem. Our speed-up estimates therefore give a reasonable expectation for others hoping to apply QDMC to a related planetary dynamics problem. It is also important to note that, given our computational resources, we would not have been able to estimate the Mercury instability event statistics as accurately as we did using direct numerical simulation. For problems like this, applying QDMC is essential even if non-negligible up-front work is required.

We have demonstrated the utility of QDMC here on an efficient N-body scheme that may not be accurate when Mercury's eccentricity becomes large \citep{zeebe2015dynamic}. In future work it would be possible to apply QDMC to the more expensive numerical scheme used by \citet{zeebe2015dynamic}. Alternatively, QDMC could be applied to a scheme like that used in this paper, and then the scheme of \citet{zeebe2015dynamic} could be switched to once Mercury's eccentricity became large.

As an alternative to QDMC,
there may be astronomical problems for which extreme value theory \citep{de2006extreme} provides an efficient approach for computing rare event probabilities.
However, extreme value theory rests on statistical assumptions that are not always satisfied \citep{naess2008monte}.
The assumptions of stationarity and of a mean upcrossing function of generalized Gumbel form are not valid for Mercury's eccentricity data (Fig. 2),
and more generally the assumptions behind extreme value theory are invalid or difficult to validate for many problems in the Earth and planetary sciences \citep{stein2020some}. In contrast, QDMC provides unbiased rare event statistics for any stochastic simulation model, making it a broadly attractive approach for computing rare probabilities.

QDMC could be used to assess the probability that a near-Earth asteroid impacts Earth, which could improve estimates of their value on the Torino Impact Hazard Scale \citep{binzel2000torino}. Additionally, a different rare event scheme such as action minimization  \citep{E2004mam,plotkin2019maximizing,woillez2020instantons} could be used to calculate the minimum impulse necessary to prevent an asteroid from colliding with Earth as long as a numerical scheme with a linear adjoint is available.

An overdensity is apparent when Mercury's eccentricity is $\sim$0.32 in Fig.~\ref{fig:trajectories-gr-dns}. This overdensity is apparent for running means of $\mathcal{O}(10^{7})$~yr, but not in the high-frequency output. The overdensity could be related to the overlap of the $g_5$ and $s_2$ resonance modes, which occurs at roughly the same eccentricity of Mercury \citep{lithwick2014secular}. The explanation for the overdensity might be an interesting problem for future research. 

In Section \ref{sec:methods:reaction_coordinate}, we demonstrated that the range of values attained by Mercury and Venus's MOID over a timescale of 400~Myr is strongly correlated with the probability that Mercury will undergo an instability over the ensuing 400~Myr. We found that this reaction coordinate was successful for our purposes since QDMC achieves a large speed-up as compared to direct numerical simulation.
However, it is possible that we could improve the efficiency further by finding an improved reaction coordinate, especially using the large sample of simulations we produced using QDMC in which Mercury's orbit becomes unstable. 
The underlying dynamical mechanisms leading to Mercury's potential instability have been explored by a number of authors \citep{LithwickWu2011secular_chaos,Boue2012,lithwick2014secular,batygin2015chaotic}.
This work has established the role of various secular resonances in driving chaotic evolution and Mercury's eventual eccentricity excitation \citep[see also][]{Laskar1990,LaskarQuinnTremaine1992,SussmanWisdom1992}. 
A reaction coordinate that measures the influence of these resonances, e.g., through frequency analysis of the system's secular modes \citep{MogaveroLaskar2021},  could potentially provide  improved algorithm performance.
More generally, improving the reaction coordinate is a rich problem for future research with many potential angles of attack.

Over extremely short time horizons ($< 1$~Gyr),
Mercury's instability probability may depend sensitively on the details of the simulation model.
Unresolved physics (e.g., asteroids, satellites of planets, Pluto, Trans-Neptunian objects, Oort Cloud objects, and other stars passing near the Solar System), the choice of integrator time-step, and the details of the noise model 
may
have a significant impact on calculated instability probabilities. 
Even rounding error can cause a measurable deviation in orbital paths over 
such short timescales.
However, our
working hypothesis, and that of other authors in the field \citep{laskar1994large,laskar2008chaotic,batygin2008dynamical,laskar2009existence,zeebe2015highly,woillez2017long}, is that Mercury instability events on time horizons of 2.4 Gyr, 2.8 Gyr, and 3.2 Gyr are not so unlikely that these model details are essential.   Testing this hypothesis on shorter time horizons is an outstanding, but computationally expensive task.

\section{Conclusion} \label{sec:conclusion}

We have shown that Quantile Diffusion Monte Carlo (QDMC) provides unbiased estimates of the probability of rare events in an N-body planetary dynamics code at greatly reduced computational cost. 
More specifically, using QDMC we were able to estimate the probability that Mercury's orbit becomes unstable $\sim$2 billion years in the future as $\mathcal{O}(10^{-4})$ at a computational cost up to 100 times lower than direct numerical simulation. QDMC is easy to implement and could be applied to many planetary dynamics problems involving rare events. 

\begin{acknowledgments}
We thank Edwin Kite for suggesting we investigate Mercury instability events using QDMC. We thank Mark Hammond for help setting up a Conda environment and iPython notebook. We thank Garett Brown and Hanno Rein for specifying parameter choices for the Rebound simulations. We thank Richard Zeebe for sharing his simulation data with us. We thank Nora Bailey, Jade Checlair, Dan Fabrycky, Gregory Gilbert, Ben Hayworth, Xuan Ji, Tad Komacek, Eric Vanden-Eijnden, and Huanzhou Yang for discussions and comments. We thank Nora Bailey, Konstantin Batygin, Freddy Bouchet, Garett Brown, Dan Fabrycky, Jacques Laskar, Hanno Rein, Richard Zeebe, and an anonymous reviewer for feedback on an early version of this paper.  This work was completed with resources provided by the University of Chicago Research Computing Center. This work was supported by the NASA Astrobiology Program grant No. 80NSSC18K0829 and benefited from participation in the NASA Nexus for Exoplanet Systems Science research coordination network. RJW was supported by New York University's Dean's Dissertation Fellowship and by the National Science Foundation through award DMS-1646339. JW acknowledges support from the Advanced Scientific Computing Research Program within the DOE Office of Science through award DE-SC0020427. SH gratefully acknowledges the CfA Fellowship.
\end{acknowledgments}


\begin{thebibliography}{}
\expandafter\ifx\csname natexlab\endcsname\relax\def\natexlab#1{#1}\fi
\providecommand{\url}[1]{\href{#1}{#1}}
\providecommand{\dodoi}[1]{doi:~\href{http://doi.org/#1}{\nolinkurl{#1}}}
\providecommand{\doeprint}[1]{\href{http://ascl.net/#1}{\nolinkurl{http://ascl.net/#1}}}
\providecommand{\doarXiv}[1]{\href{https://arxiv.org/abs/#1}{\nolinkurl{https://arxiv.org/abs/#1}}}

\bibitem[{Allen {et~al.}(2006)Allen, Frenkel, \& Rein~ten
  Wolde}]{AllenFrenkel:2006:FFS}
Allen, R.~J., Frenkel, D., \& Rein~ten Wolde, P. 2006, J. Chem. Phys., 124,
  024102

\bibitem[{Antoszewski {et~al.}(2020)Antoszewski, Feng, Vani, Thiede, Hong,
  Weare, Tokmakoff, \& Dinner}]{Antoszewski2020insulin}
Antoszewski, A., Feng, C.-J., Vani, B.~P., {et~al.} 2020, The Journal of
  Physical Chemistry B, 124, 5571, \dodoi{10.1021/acs.jpcb.0c03521}

\bibitem[{Batygin \& Laughlin(2008)}]{batygin2008dynamical}
Batygin, K., \& Laughlin, G. 2008, The Astrophysical Journal, 683, 1207

\bibitem[{Batygin {et~al.}(2015)Batygin, Morbidelli, \&
  Holman}]{batygin2015chaotic}
Batygin, K., Morbidelli, A., \& Holman, M.~J. 2015, The Astrophysical Journal,
  799, 120

\bibitem[{Binzel(2000)}]{binzel2000torino}
Binzel, R.~P. 2000, Planetary and Space Science, 48, 297

\bibitem[{{Bou{\'e}} {et~al.}(2012){Bou{\'e}}, {Laskar}, \&
  {Farago}}]{Boue2012}
{Bou{\'e}}, G., {Laskar}, J., \& {Farago}, F. 2012, \aap, 548, A43,
  \dodoi{10.1051/0004-6361/201219991}

\bibitem[{Brown \& Rein(2020)}]{brown2020repository}
Brown, G., \& Rein, H. 2020, Research Notes of the AAS, 4, 221

\bibitem[{C{\'e}rou \& Guyader(2007)}]{CerouGuyader:2007:AMS}
C{\'e}rou, F., \& Guyader, A. 2007, Stochastic Analysis and Applications, 25,
  417, \dodoi{10.1080/07362990601139628}

\bibitem[{De~Haan {et~al.}(2006)De~Haan, Ferreira, \& Ferreira}]{de2006extreme}
De~Haan, L., Ferreira, A., \& Ferreira, A. 2006, Extreme value theory: an
  introduction, Vol.~21 (Springer)

\bibitem[{Dean \& Dupuis(2009)}]{DeanDupuis:2009:splitting}
Dean, T., \& Dupuis, P. 2009, Stochastic Processes and their Applications, 119,
  562587

\bibitem[{Dean \& Dupuis(2011)}]{DeanDupuis:2011:AnnOpRes}
---. 2011, Annals of Operations Research, 189, 63.
\newblock \url{http://dx.doi.org/10.1007/s10479-009-0664-7}

\bibitem[{Deville \& Tille(1998)}]{deville1998unequal}
Deville, J.-C., \& Tille, Y. 1998, Biometrika, 85, 89

\bibitem[{Dupuis \& Wang(2004)}]{DupuisWang:2004:Stochastics}
Dupuis, P., \& Wang, H. 2004, Stochastics, 76, 481

\bibitem[{E {et~al.}(2004)E, Ren, \& Vanden-Eijnden}]{E2004mam}
E, W., Ren, W., \& Vanden-Eijnden, E. 2004, Communications on Pure and Applied
  Mathematics, 57, 637, \dodoi{https://doi.org/10.1002/cpa.20005}

\bibitem[{Giardina {et~al.}(2006)Giardina, Kurchan, \&
  Peliti}]{giardina2006direct}
Giardina, C., Kurchan, J., \& Peliti, L. 2006, Physical review letters, 96,
  120603

\bibitem[{Grassberger(1997)}]{Grassberger:1997:splitting}
Grassberger, P. 1997, Phys. Rev. E, 56, 3682, \dodoi{10.1103/PhysRevE.56.3682}

\bibitem[{Greene {et~al.}(2021)Greene, Webber, Berkelbach, \&
  Weare}]{greene2021approximating}
Greene, S.~M., Webber, R.~J., Berkelbach, T.~C., \& Weare, J. 2021, arXiv
  preprint arXiv:2103.12109

\bibitem[{Guttenberg {et~al.}(2012)Guttenberg, Dinner, \&
  Weare}]{GuttenbergDinner:2012:STPS}
Guttenberg, N., Dinner, A.~R., \& Weare, J. 2012, The Journal of Chemical
  Physics, 136, \dodoi{http://dx.doi.org/10.1063/1.4724301}

\bibitem[{Haraszti \& Townsend(1999)}]{HarasztiTownsend:1999:DPR}
Haraszti, Z., \& Townsend, J.~K. 1999, ACM Trans. Model. Comput. Simul., 9,
  105, \dodoi{10.1145/333296.333349}

\bibitem[{Huber \& Kim(1996)}]{HUBER1996we}
Huber, G., \& Kim, S. 1996, Biophysical Journal, 70, 97,
  \dodoi{https://doi.org/10.1016/S0006-3495(96)79552-8}

\bibitem[{Kahn \& Harris(1951)}]{KahnHarris:1951:splitting}
Kahn, H., \& Harris, T. 1951, Natl. Bureau Stand. Appl. Math. Ser., 12, 27

\bibitem[{Laskar(1989)}]{laskar1989numerical}
Laskar, J. 1989, Nature, 338, 237

\bibitem[{{Laskar}(1990)}]{Laskar1990}
{Laskar}, J. 1990, \icarus, 88, 266, \dodoi{10.1016/0019-1035(90)90084-M}

\bibitem[{Laskar(1994)}]{laskar1994large}
Laskar, J. 1994, Astronomy and Astrophysics, 287, L9

\bibitem[{Laskar(2008)}]{laskar2008chaotic}
---. 2008, Icarus, 196, 1

\bibitem[{Laskar \& Gastineau(2009)}]{laskar2009existence}
Laskar, J., \& Gastineau, M. 2009, Nature, 459, 817

\bibitem[{{Laskar} {et~al.}(1992){Laskar}, {Quinn}, \&
  {Tremaine}}]{LaskarQuinnTremaine1992}
{Laskar}, J., {Quinn}, T., \& {Tremaine}, S. 1992, \icarus, 95, 148,
  \dodoi{10.1016/0019-1035(92)90196-E}

\bibitem[{{Lithwick} \& {Wu}(2011)}]{LithwickWu2011secular_chaos}
{Lithwick}, Y., \& {Wu}, Y. 2011, \apj, 739, 31,
  \dodoi{10.1088/0004-637X/739/1/31}

\bibitem[{Lithwick \& Wu(2014)}]{lithwick2014secular}
Lithwick, Y., \& Wu, Y. 2014, Proceedings of the National Academy of Sciences,
  111, 12610

\bibitem[{Liu(2008)}]{liu2008monte}
Liu, J.~S. 2008, Monte Carlo strategies in scientific computing (Springer
  Science \& Business Media)

\bibitem[{{Mogavero} \& {Laskar}(2021)}]{MogaveroLaskar2021}
{Mogavero}, F., \& {Laskar}, J. 2021, arXiv e-prints, arXiv:2105.14976.
\newblock \doarXiv{2105.14976}

\bibitem[{Moral \& Garnier(2005)}]{DelMoral2005rare}
Moral, P.~D., \& Garnier, J. 2005, The Annals of Applied Probability, 15, 2496
  , \dodoi{10.1214/105051605000000566}

\bibitem[{Naess \& Gaidai(2008)}]{naess2008monte}
Naess, A., \& Gaidai, O. 2008, Journal of Engineering Mechanics, 134, 628

\bibitem[{Pedregosa {et~al.}(2011)Pedregosa, Varoquaux, Gramfort, Michel,
  Thirion, Grisel, Blondel, Prettenhofer, Weiss, Dubourg, Vanderplas, Passos,
  Cournapeau, Brucher, Perrot, \& Duchesnay}]{scikit-learn}
Pedregosa, F., Varoquaux, G., Gramfort, A., {et~al.} 2011, Journal of Machine
  Learning Research, 12, 2825

\bibitem[{Plotkin {et~al.}(2019)Plotkin, Webber, O'Neill, Weare, \&
  Abbot}]{plotkin2019maximizing}
Plotkin, D.~A., Webber, R.~J., O'Neill, M.~E., Weare, J., \& Abbot, D.~S. 2019,
  Journal of Advances in Modeling Earth Systems, 11, 863

\bibitem[{Ragone \& Bouchet(2021)}]{ragone2021rare}
Ragone, F., \& Bouchet, F. 2021, Geophysical Research Letters, 48,
  e2020GL091197

\bibitem[{Ragone {et~al.}(2018)Ragone, Wouters, \&
  Bouchet}]{Ragone2018heatwaves}
Ragone, F., Wouters, J., \& Bouchet, F. 2018, Proceedings of the National
  Academy of Sciences, 115, 24, \dodoi{10.1073/pnas.1712645115}

\bibitem[{Rein \& Liu(2012)}]{rein2012rebound}
Rein, H., \& Liu, S.-F. 2012, Astronomy \& Astrophysics, 537, A128

\bibitem[{{Rein} {et~al.}(2019){Rein}, {Tamayo}, \& {Brown}}]{Rein2019WHCKL}
{Rein}, H., {Tamayo}, D., \& {Brown}, G. 2019, \mnras, 489, 4632,
  \dodoi{10.1093/mnras/stz2503}

\bibitem[{Rosenbluth \& Rosenbluth(1955)}]{RosenbluthRosenbluth:1955:SIS}
Rosenbluth, M., \& Rosenbluth, A. 1955, J. Chem. Phys., 23, 356

\bibitem[{Stein(2020)}]{stein2020some}
Stein, M.~L. 2020, Statistical Science, 35, 31

\bibitem[{{Sussman} \& {Wisdom}(1992)}]{SussmanWisdom1992}
{Sussman}, G.~J., \& {Wisdom}, J. 1992, Science, 257, 56,
  \dodoi{10.1126/science.257.5066.56}

\bibitem[{{Tamayo} {et~al.}(2020){Tamayo}, {Rein}, {Shi}, \&
  {Hernandez}}]{Tamayo2020reboundx}
{Tamayo}, D., {Rein}, H., {Shi}, P., \& {Hernandez}, D.~M. 2020, \mnras, 491,
  2885, \dodoi{10.1093/mnras/stz2870}

\bibitem[{Tibshirani(1996)}]{tibshirani1996regression}
Tibshirani, R. 1996, Journal of the Royal Statistical Society: Series B
  (Methodological), 58, 267

\bibitem[{Vanden-Eijnden \& Weare(2012)}]{VandenEijndenWeare:2012:RareEvent}
Vanden-Eijnden, E., \& Weare, J. 2012, Communications on Pure and Applied
  Mathematics, 65, 1770, \dodoi{10.1002/cpa.21428}

\bibitem[{Villen-Altamirano \&
  Villen-Altamirano(1991)}]{VillenAltamiranoVillenAltamirano:1991:RESTART}
Villen-Altamirano, M., \& Villen-Altamirano, J. 1991, Proc. of the 13th
  international teletraffic congress, queuing, performance and control in ATM,
  9, 71

\bibitem[{Warmflash {et~al.}(2007)Warmflash, Bhimalapuram, \&
  Dinner}]{Warmflash:2007:NEUS}
Warmflash, A., Bhimalapuram, P., \& Dinner, A.~R. 2007, J. Chem. Phys., 127,
  154112

\bibitem[{Webber(2022)}]{webber2022variance}
Webber, R.~J. 2022, Variance estimation for splitting and killing schemes,
  European Nonlinear Dynamics Conference 2022.
\newblock
  \url{https://rwebber.people.caltech.edu/documents/20116/Webber__Weare_2022.pdf}

\bibitem[{Webber {et~al.}(2020)Webber, Aristoff, \&
  Simpson}]{webber2020splitting}
Webber, R.~J., Aristoff, D., \& Simpson, G. 2020, A splitting method to reduce
  MCMC variance.
\newblock \doarXiv{2011.13899}

\bibitem[{Webber {et~al.}(2019)Webber, Plotkin, O’Neill, Abbot, \&
  Weare}]{webber2019practical}
Webber, R.~J., Plotkin, D.~A., O’Neill, M.~E., Abbot, D.~S., \& Weare, J.
  2019, Chaos: An Interdisciplinary Journal of Nonlinear Science, 29, 053109

\bibitem[{{Wisdom}(2006)}]{Wisdom2006}
{Wisdom}, J. 2006, \aj, 131, 2294, \dodoi{10.1086/500829}

\bibitem[{Wisdom(2015)}]{wisdom2015resolving}
Wisdom, J. 2015, The Astronomical Journal, 150, 127

\bibitem[{{Wisdom} \& {Holman}(1991)}]{WH1991}
{Wisdom}, J., \& {Holman}, M. 1991, \aj, 102, 1528, \dodoi{10.1086/115978}

\bibitem[{{Wisdom} {et~al.}(1996){Wisdom}, {Holman}, \& {Touma}}]{WHT1996}
{Wisdom}, J., {Holman}, M., \& {Touma}, J. 1996, Fields Institute
  Communications, 10, 217

\bibitem[{Woillez \& Bouchet(2017)}]{woillez2017long}
Woillez, E., \& Bouchet, F. 2017, Astronomy \& Astrophysics, 607, A62

\bibitem[{Woillez \& Bouchet(2020)}]{woillez2020instantons}
---. 2020, Physical Review Letters, 125, 021101

\bibitem[{Wolf \& Toon(2015)}]{wolf2015evolution}
Wolf, E., \& Toon, O. 2015, Journal of Geophysical Research: Atmospheres, 120,
  5775

\bibitem[{Zeebe(2015{\natexlab{a}})}]{zeebe2015highly}
Zeebe, R.~E. 2015{\natexlab{a}}, The Astrophysical Journal, 811, 9

\bibitem[{Zeebe(2015{\natexlab{b}})}]{zeebe2015dynamic}
---. 2015{\natexlab{b}}, The Astrophysical Journal, 798, 8

\end{thebibliography}

\appendix

\section{Implementation details for QDMC} \label{app:implementation}

Here, we explain in more detail our procedure for randomly choosing the number of children $N_t^i$ for a simulation $X_t^i$
and assigning weights to the children.
For simplicity, we assume that the unstable trajectories have been set aside for analysis later,
and the indices are sorted so that $i = 1$ corresponds to the lowest reaction coordinate value and $i = i_{\max}$ corresponds to the highest reaction coordinate value.
Then, we perform the following three steps.

\emph{Step one}. 
We partition the indices $i = 1, \ldots, i_{\max}$ into sets \begin{align}
& S^1 = \left\{L^1, L^1 + 1, \ldots, U^1\right\}, \\
& S^2 = \left\{L^2, L^2 + 1, \ldots, U^2\right\}, \\
& \ldots, \\
& S^{\max} = \left\{L^{\max}, L^{\max} + 1, \ldots, U^{\max}\right\},
\end{align}
where the indices $\left(L^i, U^i\right)$ are determined by setting $U^0 = 0$
and setting
\begin{equation}
    L^i = U^{i-1} + 1,
    \qquad U^i = \min\left\{j \geq L^i \colon \,
    \sum_{k = L^i}^j \mathbb{E} N^k_t \geq 1\right\},
    \qquad i = 1, 2, \ldots, i_{\max}.
\end{equation}

\emph{Step two}. 
For each set $S^i$, we choose a random number
\begin{equation}
\label{eq:dist}
N_t\left(S^i\right) = 
\begin{cases}
\left\lfloor \sum_{j \in S^i} \mathbb{E} N_t^j \right\rfloor + 1, & \text{with probability }
\sum_{j \in S^i} \mathbb{E} N_t^j
- \left\lfloor \sum_{j \in S^i} \mathbb{E} N_t^j \right\rfloor,
\\[3ex]
\left\lfloor \sum_{j \in S^i} \mathbb{E} N_t^j \right\rfloor,
& \text{otherwise}.
\end{cases}
\end{equation}
Instead of choosing the numbers $N_t\left(S^i\right)$ independently,
we use pivotal sampling \citep{deville1998unequal} to ensure the random numbers $N_t\left(S^i\right)$ satisfy \eqref{eq:dist} and also
$\sum_{i=1}^{i_{\max}} N_t\left(S^i\right) = N$.

\emph{Step three}.
For each index $j \in S^i$, we choose a random number
\begin{equation}
\label{eq:dist2}
N_t^j = 
\begin{cases}
\left\lfloor \frac{N_t\left(S^i\right) w_t^j}{\sum_{k \in S^i} w_t^k} \right\rfloor + 1, & \text{with probability }
\frac{N_t\left(S^i\right) w_t^j}{\sum_{k \in S^i} w_t^k}
- \left\lfloor \frac{N_t\left(S^i\right) w_t^j}{\sum_{k \in S^i} w_t^k} \right\rfloor,
\\[3ex]
\left\lfloor \frac{N_t\left(S^i\right) w_t^j}{\sum_{k \in S^i} w_t^k} \right\rfloor,
& \text{otherwise},
\end{cases}
\end{equation}
We use pivotal sampling to ensure the numbers $N_t^j$ satisfy \eqref{eq:dist2} and also
$\sum_{k \in S^i} N_t^k = N_t\left(S^i\right)$,
and we assign the children of $X_t^j$ updated weights
$\sum_{k \in S^i} w_t^k \slash N_t\left(S^i\right)$.

For more discussion motivating these choices, see \cite{webber2020splitting} and \cite{greene2021approximating}.

\end{document}